\documentclass[aps,prl,twocolumn,showpacs]{revtex4-1}
\usepackage{bm}
\usepackage{mathrsfs}
\usepackage{amsmath}
\usepackage{amssymb}
\usepackage{graphicx}
\usepackage{amsfonts}
\usepackage{amsthm}
\usepackage{color}
\usepackage{dcolumn}
\usepackage{txfonts}
\usepackage{graphicx,amsmath,units,xspace,amssymb}

\begin{document}
\title{Probing the role of mobility in the collective motion of non-equilibrium systems}
\author{Hongchuan Shen$^1$, Peng Tan$^{2,1 *}$, and Lei Xu$^{1}$}
\email{xulei@phy.cuhk.edu.hk or tanpeng@fudan.edu.cn}
\affiliation{$^1$Department of Physics, The Chinese University of Hong Kong, Hong Kong, China\\
$^2$ State Key Laboratory of Surface Physics, Department of Physics, Fudan University, Shanghai 200433, China}
\date{\today}

\begin{abstract}
 {By systematically varying the mobility of self-propelled particles in a two-dimensional (2D) lattice, we experimentally study the influence of particle mobility on system's collective motion. Our system is intrinsically non-equilibrium due to the lack of energy equipartition. By constructing the covariance matrix of spatial fluctuations and solving for its eigenmodes, we obtain the collective motions of the system with various magnitudes. Interestingly, our structurally ordered non-equilibrium system exhibits almost identical properties as disordered glassy systems under thermal equilibrium: the modes with large overall motions are spatially correlated and quasilocalized while the modes with small collective motions are highly localized, resembling the low- and high-frequency modes in glass. More surprisingly, a peak similar to the boson peak forms in our non-equilibrium system as the number of mobile particles increases, revealing the possible origin of the boson peak from a dynamic aspect. We further illustrate that the spatially-correlated large-movement modes can be produced by the cooperation of highly-active particles above a threshold fraction, while the localized small-movement modes can be created by adding individual inactive particles. Our study clarifies the role of mobility in collective motions, and further suggests a promising possibility of extending the powerful mode analysis approach to non-equilibrium systems.}
\normalsize
\end{abstract}

\maketitle

Studying the collective motions or vibrational modes in solids is an important topic for condensed matter physics, which plays an essential role in understanding the heat capacity, sound propagation and thermal conductivity of solids. The powerful mode-analysis approach beautifully extracts the collective behaviors of the entire system from the motions of numerous individual particles. From these modes, deep insights of the system at different length and time scales can be obtained. In particular, this analysis has been widely applied to equilibrium systems, where `equilibrium' means a stable or metastable state that satisfies energy equipartition. However, the more general situation of non-equilibrium system is largely unexplored. Using active-matter systems, we now tackle this important question at single-particle level.

In equilibrium systems, energy equipartition simplifies the problem by ensuring every particle and mode to have the same energy, $\frac{1}{2}k_BT$. In crystals, the vibrational modes are plane waves and excellently described by the Debye model. However, due to the structural disorders that break the translational invariance, the modes in disordered glassy systems are much more complex and interesting~\cite{chumakov2004collective,wyart2005effects,shintani2008universal,monaco2009breakdown,kaya2010normal,huisman2011internal,green2011density,keppens1998localized,pohl2002low,schober2004vibrations,shintani2008universal,yunker2011phonon,zhang2009thermal,zhang2011prl,wang2013soft,zhao2011prl,wang2015prl}. In particular, at low frequencies strong motions tend to concentrate at the defective soft spots while the overall background is still plane-wave like~\cite{ghosh2010density,kaya2010normal,xu2010anharmonic,tan2012understanding,gratale2013phonons,chen2013phonons}, producing quasilocalized modes that play a crucial role in system rearrangements and relaxation~\cite{chen2011measurement,ghosh2011connecting,chen2013phonons}. The number density of these low-frequency modes also significantly exceeds the crystalline counterpart and forms the so called `boson peak', whose origin is still under active debate\cite{schober2004vibrations,Klinger1983,Schirmacher1998,Grigera2003,PNAS2009,bosonpeak2011}. As the frequency increases to the intermediate range, the modes become extended and uncorrelated. Once the high-frequency regime is reached, significant motions will only concentrate at very few rigid sites which produces highly-localized modes ~\cite{silbert2009normal,gratale2013phonons}.

\begin{figure}
  \centering
  \includegraphics[width=3.2in]{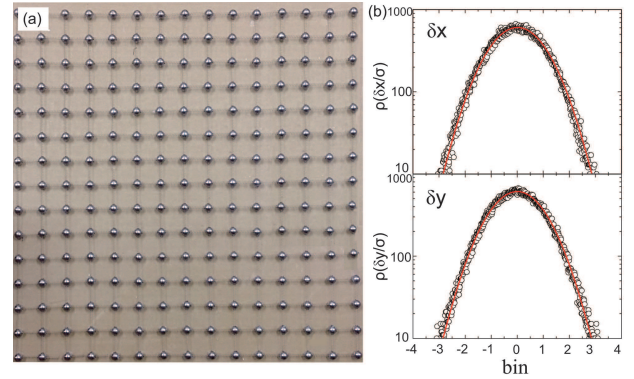}\\
  \caption{\label{fig.1} (a) The image of our system. Identical metal spheres driven by independent motors are connected by springs to form a 2D square lattice. (b) The spatial fluctuations in both $x$ and $y$ directions for all particles and frames. The data for each particle are renormalized by the local variance $\sigma$. The red line is the standard Gaussian function.}
\end{figure}

However, in non-equilibrium systems the important condition of energy equipartition breaks down. As a result, individual particles may have distinct kinetic energies or mobilities and thus the heterogeneity or `disorder' from a purely dynamic aspect can naturally arise. How does this special type of dynamic disorder affect the system? Is it similar to or different from the structural disorder and can the two be understood in a unified picture? This fundamental issue underlies numerous non-equilibrium systems that do not satisfy energy equipartition. Moreover, clarifying this issue is also crucial for the important field of active-matter systems currently under intensive research, such as the cooperative behaviors of bacteria, birds or fish within a large group, the collective motion of self-driven colloids, and the global movements and patterns of granular materials under external excitation \cite{umbanhowar1996localized,angelini2011glass,bialke2012crystallization,henkes2011active,loi2011effective,redner2013structure,tailleur2008statistical,brito2010elementary,henkes2012extracting}. However, due to the difficulty in controlling and adjusting the kinetic energy or mobility of every single particle, this fundamental puzzle remains an outstanding open question.

\begin{figure*}[!t]
  \centering
  \includegraphics[width=6.8in]{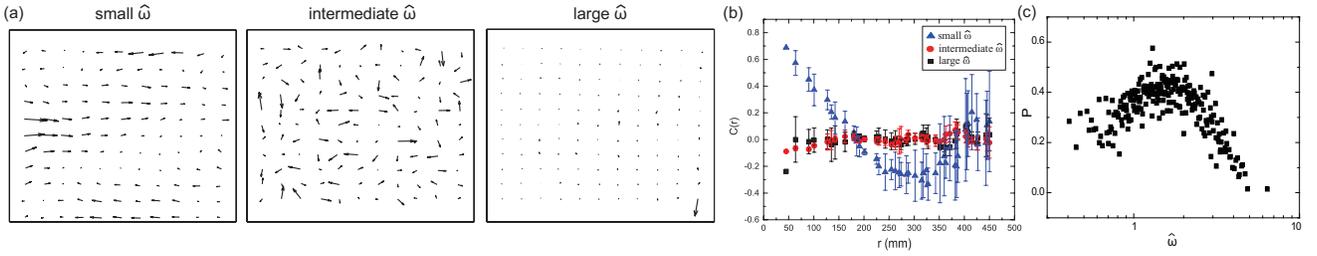}\\
  \caption{(a) Three typical eigenmodes at small, intermediate and large $\hat\omega$. (b) The spatial correlation function in direction for the three typical modes. (c) The participation ratio $P$ of all the modes.}\label{fig 2}
\end{figure*}

Using self-propelled active particles confined in a 2D square lattice, we systematically address this puzzle at single-particle level. Our particles are identical metal spheres with $d=13.00\pm0.01mm$ and $m=9.915\pm0.005g$(total mass of one sphere plus one motor). Each particle is connected to four nearest neighbors by identical springs ($k=2.97\pm0.16N/m$, $l_0=15.46\pm0.36mm$) and form a square lattice as shown in Fig.1a. All springs are stretched to reach the lattice constant of $45.0\pm1.5mm$ and the entire system contains $15\times15=225$ particles. To control the mobility at single-particle level, under \emph{every} particle we attach a small vibrating motor independently driven by external power input. Once turned on, motors will drive particles to move randomly around their equilibrium positions in 2D (see the supplemental movie). Since the attractive interaction between particles is harmonic, the spatial fluctuations (renormalized by local variance) obey an excellent Gaussian distribution as shown in Fig.1b. Clearly every particle has a well-defined equilibrium position, and the time-averaged deviation from it, $\langle\delta r_i\rangle=\langle \sqrt {\delta x_i^2+\delta y_i^2} \rangle$, provides a good description for the mobility of each particle. Due to the distinct motor mobilities, $\langle\delta r_i\rangle$'s at different sites have a typical dispersion above $15\%$, which is much larger than the variations in lattice constant, particle mass, and interaction potential. Therefore, our system provides an ideal platform to probe the influence of mobility heterogeneity because both the structure and the interaction are set to be the simplest situation while only mobility varies significantly.

To understand the role of this mobility disorder in the collective motion, we construct the covariance matrix of spatial fluctuations and calculate its eigenmodes \cite{tan2012understanding,henkes2012extracting,chen2010low}. This particular principal component analysis (PCA) on spatial fluctuations has great potential in extracting the system's collective movements: in equipartitioned equilibrium systems the eigenmodes are identical to the vibrational modes \cite{henkes2012extracting}; in non-equilibrium systems the eigenmodes are not the same as vibrational modes any more however they still reveal the system¡¯s collective motions at single-particle level.

More specifically, we track the positions of all particles for 1250 frames (see SI for the influence of frame number) and construct the covariance matrix \cite{tan2012understanding,henkes2012extracting,chen2010low}: $C_{i, j}=\langle[\bm{r}_i(t)-\langle \bm{r}_i(t)\rangle][\bm{r}_j(t)-\langle \bm{r}_j(t)\rangle]\rangle$, with $i, j= 1, ... , 2N_p$ running over the x and y coordinates of all particles, and $\langle\rangle$ indicating time average over all frames. To eliminate the boundary effect, we only use the central $N_p=11\times11=121$ particles which result in $2N_p=242$ eigenmodes. In equilibrium systems, these eigenmodes are identical to the vibrational modes, with the eigenvalue $\lambda$ directly related to the vibrational frequency $\omega$: $\omega \propto 1/\sqrt\lambda$ \cite{ghosh2010density,henkes2012extracting,chen2010low}. Analogous to $\omega$, therefore, in our non-equilibrium system we define a dimensionless parameter, $\hat{\omega}\equiv \langle\overline{\delta r}\rangle/\sqrt\lambda$, which has the same $\lambda$ dependence and renormalized by the time-and-location averaged displacement $\langle\overline{\delta r}\rangle$. Due to the lack of equipartition, our eigenmodes are \emph{not} equivalent to vibrational modes anymore and $\hat\omega$ is \emph{not} the vibrational frequency, but these modes still indicate specific patterns of collective movements (see Fig.2a), following which the system can achieve the overall displacement magnitude described by $\sqrt\lambda$ or $1/\hat\omega$ \cite{brito2010elementary,henkes2012extracting}.

\begin{figure*}
  \centering
  \includegraphics[width=6.8in]{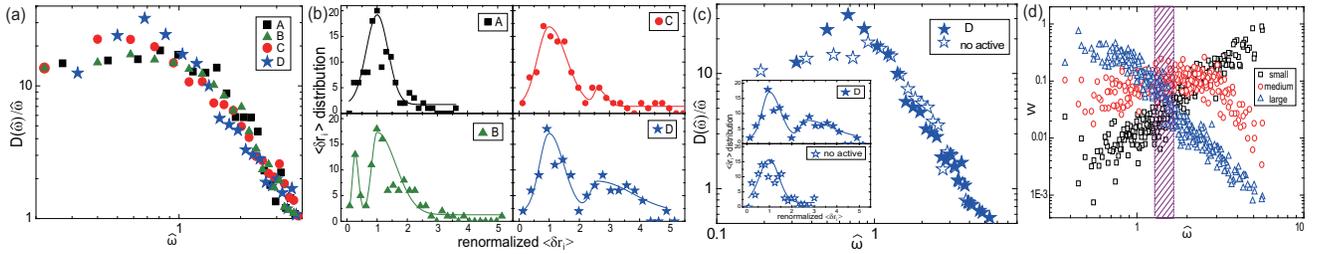}\\
  \caption{(a) The $D(\hat\omega)/\hat\omega$ spectra of four typical samples, A to D, demonstrate the formation of a low-$\hat\omega$ peak. (b) The distribution of time-averaged local displacement, $\langle\delta r_i\rangle$, for samples A to D. To achieve such distributions, the four samples are driven with different voltages: A ($2.5V$), B (normally $2.5V$ with $16.5\%$ of $2V$), C (normally $2.5V$ with $20.7\%$ of $3.2V$) and D (normally $2.5V$ with $20.7\%$ of $3.2V$ and $20.7\%$ of $3.6V$). $\langle\delta r_i\rangle$'s are renormalized to make the main peak locate at 1. (c) Tuning the active particles in sample D back to normal activities eliminates the peak. Inset shows the $\langle\delta r_i\rangle$ distribution with and without active particles. (d) The relative importance of small-, medium- and large-mobility groups throughout all $\hat\omega$. The hatched area where all groups are comparable corresponds to the peak in the participation ratio.}\label{fig 3}
\end{figure*}

We show three typical modes at small, intermediate and large $\hat\omega$ in Fig.2a: at small $\hat\omega$ polarization vectors exhibit large-scale correlations, at intermediate $\hat\omega$ they are rather random, and at high $\hat\omega$ motions localize on very few specific sites. These behaviors excellently agree with glassy systems under thermal equilibrium \cite{ghosh2010density,henkes2012extracting,chen2010low}. To quantify the spatial correlation, we plot the directional correlation function, $C(r)=\sum_{i,j=1}^{Np}\delta(r_{ij}-r)\hat{\textbf{e}}_{\hat\omega,i}\cdot\hat{\textbf{e}}_{\hat\omega,j}/\sum_{i,j=1}^{Np}\delta(r_{ij}-r)$ (here $\hat{\textbf{e}}_{\hat\omega,i}$ and $\hat{\textbf{e}}_{\hat\omega,j}$ are unit polarization vectors in mode $\hat\omega$), for the three modes in Fig.2b: apparently large-scale correlation only exists at small $\hat\omega$ which is consistent with Fig.2a.

To illustrate the distribution of motion among different sites, we plot the participation ratio, $P(\hat\omega)=(\sum_i |\textbf{e}_{\hat\omega,i}|^2)^2/(N_p\sum_i |\textbf{e}_{\hat\omega,i}|^4)$, in Fig.2c. A smaller $P$ indicates a more localized mode and vice versa. Clearly, $P$ starts relatively small at low $\hat\omega$, increases continuously to reach a peak at intermediate $\hat\omega$, and then decreases to rather small values at high $\hat\omega$. This trend of quasilocalized to extended then to localized in the $P$ spectrum again matches the equilibrium glassy systems very well. More careful inspection further reveals that at low $\hat\omega$ large motions concentrate on large-mobility particles, while at high $\hat\omega$ they localize at small-mobility particles. Comparing with previous studies on glassy systems, our large-mobility particles naturally correspond to the defective soft spots, and the small-mobility ones are analogous to the rigid spots\cite{gratale2013phonons}. It suggests that the mobility disorder makes a similar influence as the structure and interaction disorder, raising the intriguing possibility of a unified understanding.

We illustrate the mode distribution of different-mobility samples with the spectrum analogous to the reduced density of states, $D(\hat\omega)/\hat\omega$, in Fig.3a. To clarify the influence of particle mobility on the spectrum, we prepare sample A with a regular Gaussian $\langle\delta r_i\rangle$ distribution, sample B with extra small-mobility inactive particles, and samples C and D with increasing amount of large-mobility active particles, as shown in Fig.3b. Interestingly, in sample A, $D(\hat\omega)/\hat\omega$ is quite flat at low $\hat\omega$, which resembles the 2D crystal described by Debye model. Similarly, in sample B the low-$\hat\omega$ spectrum remains flat, indicating that adding inactive particles makes negligible influence. However, as more and more active particles are introduced into samples C and D, a low-$\hat\omega$ peak develops in an approach resembling the boson peak formation in glass. These data strongly suggest that the particles with large mobilities can produce extra low-$\hat\omega$ modes and cause boson peak; while the inactive ones have no such effect.

To completely confirm this conclusion, we then tune the active particles in samples C and D back to normal activities and the peak disappears, as shown in Fig.3c and Fig.SI-3. This result unambiguously verifies that the low-$\hat\omega$ peak is indeed caused by large-mobility particles. Comparing with the boson peak formation due to defects in glass, our active particles naturally correspond to the defective \emph{soft} spots which are considered as the origin of boson peak \cite{shintani2008universal}. With the mobility analysis, our study illustrates the origin of boson peak from the aspect of mobility, and suggests large mobility as the possible origin mechanism for boson peak.

\begin{figure*}
  \centering
  \includegraphics[width=6.8in]{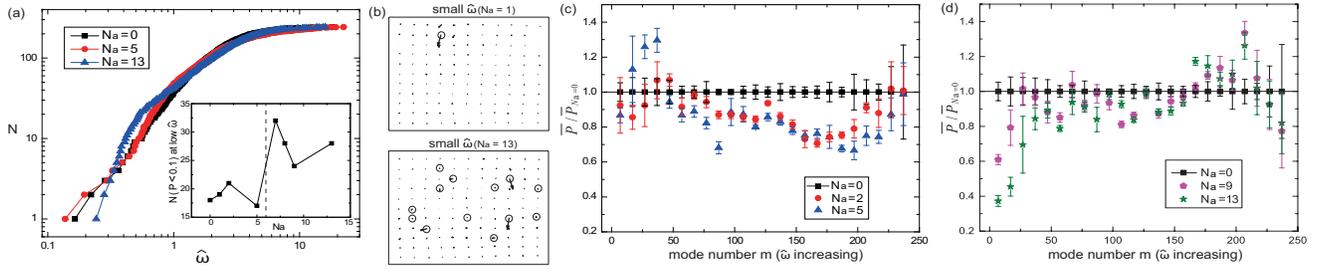}\\
  \caption{(a) The variation in cumulative number of modes $N(\hat\omega)$ as $N_a$ active particles are added. Inset: the number of low-$\hat\omega$ quasilocalized modes ($P<0.1$) versus $N_a$ increases abruptly around the threshold value of $N_a=6$ (i.e., $5\%$). (b) Large motions occur at the active-particle sites (label by circles) in the low-$\hat\omega$ modes. (c) and (d) compare the participation ratio in systems with and without active particles. (c) shows the situation of $N_a<6$ and (d) demonstrates the situation of $N_a>6$.}\label{fig 5}
\end{figure*}

Clearly the active particles can significantly affect the low-$\hat\omega$ modes. To obtain a complete picture, we further probe the importance of different-mobility particles throughout the entire $\hat\omega$ range. We pick three typical groups of particles with small, medium and large mobilities respectively, and measure their relative importance in all eigenmodes. Each group contains 11 particles ($9\%$ of total $N_p$) and the data are shown in Fig.3d: as $\hat\omega$ increases, the importance of large-mobility group changes from dominant to negligible; while the small-mobility group exhibits a completely opposite trend. Note that the $9\%$ of large- and small-mobility particles account for almost $100\%$ of the motion in the low- and high-$\hat\omega$ modes respectively. By contrast, the medium-mobility group shows negligible importance at both low- and high-$\hat\omega$ regions, and only exhibits comparable weight as the other two groups around the intermediate $\hat\omega$ indicated by a hatched area. This hatched region corresponds excellently to the peak position of the participation ratio in Fig.2c, as we naturally expect from the definition of extended modes.

To obtain an exact understanding on the role of particle mobility and further achieve desirable manipulation on the mode spectrum, we systematically introduce highly-active or inactive particles into our system. Active particles all have mobilities above 240\% of normal value; and inactive particles have mobilities below 45\% of normal value. First we replace normal particles with $N_a$ highly-active particles and plot the cumulative number of modes, $N(\hat\omega)$, in Fig.4a. Significant change only occurs at low $\hat\omega$. More specifically, the curve with small amount of active particles ($N_a=5$, about $4\%$) does not show a noticeable deviation from the original $N_a=0$ curve; while the curve with $N_a=13$ (about $11\%$) shows a much sharper increase at the beginning, indicating the appearance of more low-$\hat\omega$ eigenmodes.

To get a quantitative understanding, we measure the number of low-$\hat\omega$ quasilocalized modes ($P<0.1$) in systems with different $N_a$ and plot it in the inset: an abrupt jump appears around $N_a=6$ (i.e., $5\%$) as indicated by the dashed line. This threshold behavior implies that the low-$\hat\omega$ spectrum can only be significantly modified by adding \emph{enough} mobile particles. Some typical low-$\hat\omega$ modes are directly visualized in Fig.4b which confirms that large motions tend to occur at the highly-active particles (labeled by circles). Although in the $N_a=13$ image only a fraction of active particles exhibit large motions, other active particles experience similar large motions in the neighboring modes not shown here.

We further compare the participation ratio, $P$, for systems with different $N_a$ against the original $N_a=0$ system. To reduce random fluctuations, we average $P$ over ten neighboring modes to obtain $\overline{P}$, and calculate the ratio, $\overline{P}/\overline{P}_{Na=0}$, for all the modes. The values of $\overline{P}/\overline{P}_{Na=0}$ versus the mode number $m$ (in the direction of increasing $\hat\omega$) are plotted in Fig.4c and Fig.4d. Again we find two distinct behaviors: small amount of active particles do not produce systematic variations at low $\hat\omega$ although they seem to reduce the participation ratio at intermediate and high $\hat\omega$ (see Fig.4c); however, after $N_a$ passes the threshold value, the participation ratio at low-$\hat\omega$ reduces continuously with $N_a$ (see Fig.4d). Together with the similar threshold behavior observed in Fig.4a, we conclude that the low-$\hat\omega$ spectrum can not be modified by adding one or two active particles, instead a threshold amount is required to create \emph{new} spatially-correlated low-$\hat\omega$ modes (not to just disturb existing ones but to create \emph{new} ones). Similar requirement may also hold for the generation of low-frequency modes in equilibrium glassy systems, although the exact threshold value may vary with specific conditions such as dimensionality, pressure and temperature. The surprising appearance of small $N_a$'s influence to the intermediate- and high-$\hat\omega$ modes in Fig.4c is not understood and calls for further investigation.

Similarly, we can systematically replace normal particles with inactive particles. We again plot the cumulative number of modes, $N(\hat\omega)$, for systems with $N_i$ inactive particles in Fig.5a. Because the inactive particles mainly affect large-$\hat\omega$ region, we use $1/\hat\omega$ as the x-axis which better stresses any change at large $\hat\omega$. As $N_i$ increases, large-$\hat\omega$ modes are shifted to even higher $\hat\omega$. The quantitative measurements in the inset demonstrate that the number of high-$\hat\omega$ localized modes increases continuously with $N_i$ and saturates at large $N_i$. This continuously increasing trend makes a sharp contrast to the abrupt jumping behavior for $N_a$ in the Fig.4a inset, and the saturation is possibly due to the finite size of our system. Direct visualization of the high-$\hat\omega$ localized modes are shown in Fig.5b: large motions tend to localize on the inactive-particle sites (labeled by circles) which confirms that the newly-added inactive particles are responsible for these localized modes. For their influence on participation ratio, we again use the values in the original system, $\overline{P}_{Ni=0}$, as the control and plot the ratio $\overline{P}/\overline{P}_{Ni=0}$ in Fig.5c: as $N_i$ increases, the ratio drops significantly at high $\hat\omega$ while no consistent trend is observed in the low-$\hat\omega$ region (see Fig.SI-4 for more data).

\begin{figure}[!b]
  \centering
  \includegraphics[width=3.4in]{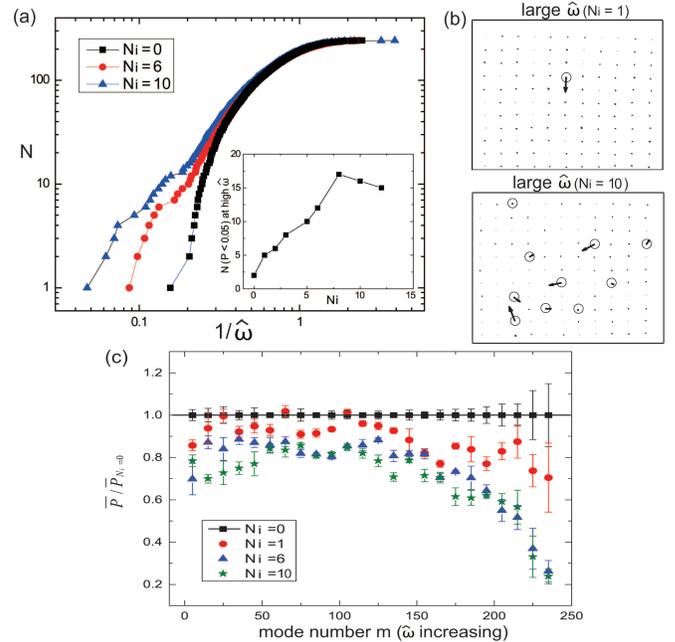}\\
  \caption{(a) The cumulative number of modes $N(1/\hat\omega)$ for different number of inactive particles. Inset, the number of high-$\hat\omega$ localized modes ($P<0.05$) versus $N_i$. (b) Motions are concentrated at inactive particles (labeled by circles) in high-$\hat\omega$ modes. (c) As $N_i$ increases, the participation ratio in the large-$\hat\omega$ region decreases significantly. }\label{fig 5}
\end{figure}

To summarize, the creation of spatially-correlated low-$\hat\omega$ modes requires the collaboration of quite a few large-mobility particles, while the generation of localized high-$\hat\omega$ modes only needs individual inactive particles. The two distinct behaviors may also hold in glassy systems, and provide a potential guidance for the manipulation of mode spectrum in glass.

In conclusion, we have studied the role of particle mobility in the collective motions of 2D systems. Our results reveal that large- and small-mobility particles can significantly affect low- and high-$\hat\omega$ modes respectively, which enables an effective manipulation on the spectrum by adding active or inactive particles. Although our system is on an ordered square lattice, the difference in particle mobility causes collective motions very similar to disordered glassy systems. Therefore the mobility disorder plays a similar role as the structural and potential disorder, which suggests a possible unification of all types of disorders. The quantities based on the time-averaged local displacement $\langle\delta r_i\rangle$ or local debye-waller factor can contain information from both the dynamic and the structural influences, and may provide a good order parameter for this unification. Actually such kind of order parameter has already demonstrated its great potential in the simulation of jammed spheres \cite{tong2014structural}.

Moreover, we make further speculations on the comparison between equilibrium and non-equilibrium systems. With the knowledge of spring constant $k$, particle mass $m$ and system size, we can theoretically obtain the range of vibrational frequencies for our lattice as $9.75$ to $34.62$Hz. By assuming that each mode roughly has the energy of $\frac{1}{2}k\langle\overline{\delta r}\rangle^2$ we can also get the frequency range from covariance matrix measurements: $\hat\omega\sqrt{\frac{k}{m}}\sim3.6$ to 372.7Hz. Note that this assumption is not strictly valid due to the lack of energy equipartition, and we just do it for the comparison with theoretical frequencies. Apparently the two results overlap reasonably well in the low frequency regime while at high frequencies the latter significantly exceeds the former, indicating that equal-partition might roughly hold for spatially-correlated low-$\hat\omega$ modes while serious break-down only occurs at high frequencies. Similar to equilibrium systems, we also find correlations between real displacements and eigenmodes, as demonstrated in the projection plot in Fig.SI-5. Our study on non-equilibrium systems suggests the possibility of extending the powerful mode-analysis approach from equilibrium systems to non-equilibrium active matter or even biological systems.

\begin{acknowledgements}
This project is supported by Hong Kong Research Grants Council under the projects of Early Career Schemes Grant No. CUHK404912 and General Research Fund Grant No. CUHK14303415; CUHK Faculty of Science under the Direct Grant No. 4053131 and Science Faculty Young Researcher Award 2014; and the National Natural Science Foundation of China No. 11504052.
\end{acknowledgements}


\end{document}